# Spectre Returns! Speculation Attacks using the Return Stack Buffer

*Esmaiel Mohammadian Koruyeh, Khaled Khasawneh,*
*Chengyu Song and Nael Abu-Ghazaleh*
*Computer Science and Engineering Department*
*University of California, Riverside*
`naelag@ucr.edu`

## Abstract

The recent Spectre attacks exploit speculative execution, a pervasively used feature of modern microprocessors, to allow the exfiltration of sensitive data across protection boundaries. In this paper, we introduce a new Spectre-class attack that we call SpectreRSB. In particular, rather than exploiting the branch predictor unit, SpectreRSB exploits the return stack buffer (RSB), a common predictor structure in modern CPUs used to predict return addresses. We show that both local attacks (within the same process such as Spectre 1) and attacks on SGX are possible by constructing proof of concept attacks. We also analyze additional types of the attack on the kernel or across address spaces and show that under some practical and widely used conditions they are possible. Importantly, none of the known defenses including Retpoline and Intel's microcode patches stop all SpectreRSB attacks. We believe that future system developers should be aware of this vulnerability and consider it in developing defenses against speculation attacks. In particular, on Core-i7 Skylake and newer processors (but not on Intel's Xeon processor line), a patch called RSB refilling is used to address a vulnerability when the RSB underfills; this defense interferes with SpectreRSB's ability to launch attacks that switch into the kernel. We recommend that this patch should be used on all machines to protect against SpectreRSB.

## 1 Introduction

Speculative execution is a microarchitectural technique used pervasively to improve the performance of all modern CPUs. Recently, it has been shown that speculatively executed instructions can leave measurable side-effects in the processor caches and other shared structures even when the speculated instructions do not commit and their direct effects are not visible. Moreover, since these instructions are speculative, normal permission checks do not take effect until the instruction is committed. The recent Spectre attack [23, 13, 31] has shown that this behavior can be exploited to expose information that is otherwise inaccessible. In the two variants of Spectre attacks, attackers either mistrain the branch predictor unit or directly pollute it to force the speculative execution of code that can enable exposure of the full memory of other processes and hypervisor.

Chen et al. demonstrated that known Spectre variants are able to expose information from SGX enclaves [3]. New variants of Spectre that utilize other triggers for speculative execution have been introduced including speculative store bypass [17].

In this paper, we introduce a new attack vector Spectre like attacks that are not prevented by deployed defenses. Specifically, the attacks exploit the Return Stack Buffer (RSB) to cause speculative execution of the payload gadget that reads and exposes sensitive information. The RSB is a processor structure used to predict return address by pushing the return address from a call instruction on an internal hardware stack (typically of size 16 entries). When the return is encountered, the processor uses the top of the RSB to predict the return address to support speculation with very high accuracy.

We show that the RSB can be easily manipulated by user code: a `call` instruction, causes a value to be pushed to the RSB, but the stack can subsequently be manipulated by the user so that the return address no longer matches the RSB. We describe the behavior of RSB in more details in Section 3.

In Section 4, we show an RSB based attack that accomplishes the equivalent of Spectre variant 1 through manipulation of the RSB instead of mistraining the branch predictor; we use this scenario to explain the principles of the attack, even though it may not be practical.The RSB is shared among hardware threads that execute on the same virtual processor enabling inter-process (or even inter-vm) pollution of the RSB. Thus, in Section 5, we develop an attack that targets a different thread or process on the same machine. In Section 6, we present a third type of SpectreRSB: an attack against an SGX compartment where a malicious OS pollutes the RSB to cause a misspeculation that exposes data outside an SGX compartment. This attack bypasses all software and microcode patches on our SGX machine. Section 7 overviews another potential attack targeting an unmatched return in the kernel code; we present this attack for completeness because it relies on a number of ingredients that are difficult to find in practice.

We show how these attacks interact with deployed defenses concluding that several practical deployments are vulnerable to SpectreRSB. Thus, we believe that SpectreRSB is as a dangerous speculation attack, that in some instances is not mitigated by the primary defenses against Spectre. It extends our understanding of the threat surface of speculation attacks, allowing future defenses to more effectively mitigate their risks. We discuss implications of the attack in Section 8.

Section 9 overviews related work, while Section 10 presents some concluding remarks.

**Disclosure:** We reported these attacks to the security team at Intel. Although we did not demonstrate attacks on AMD and ARM processors, they also use RSBs to predict return addresses. Therefore, we also reported our results to AMD and ARM.

## 2 Speculation Attacks and Defenses

Speculative execution has been an important part of computer architecture starting from the 1950s. The IBM Stretch processor implemented a predict not-taken branch predictor to avoid stalling a processor pipeline when a branch is encountered [1]. Computer architecture advanced rapidly starting in the early 1980s leading to rapid increase in the amount of speculation that is exploited with aggressive out-of-order execution. This speculation is supported by sophisticated branch predictor designs [41, 18, 35] that are highly successful in predicting both the branch direction and its target address. In particular, the number of pipeline stages in production CPUs has continued to grow to the point where modern pipelines commonly have between 15 and 25 stages. With out-of-order execution, when a branch instruction stalls (e.g., due to a cache miss on which it depends), instructions that follow the branch are continuously being issued. Thus, the speculation window where instructions are getting executed speculatively can be large, typically limited by the size of structures such as the reorder buffer, which can hold a few hundred instructions.

Speculation is designed to not affect the correctness of a program. Although branch mispredictions occur and speculative instructions can ignore execution faults (e.g., permission error for memory access) these semantics were not considered harmful as mis-speculation will eventually be detected and the erroneously executed instructions will be squashed, leaving no directly visible changes to the program state held in structures such as registers and memory. Micro-architectural structures such as caches and Translation Look-aside Buffers (TLB) are affected by speculative operations, but the contents of such structures typically only affect performance, not the correctness of a program. In fact, prior work has shown that there are beneficial prefetching side-effects to speculatively executed instructions even those that are eventually squashed [33].

### 2.1 Speculation Attacks

Spectre attacks have recently shown that the speculation behavior of modern processors can be exploited. In general, these attacks exploit four properties:

- **P1**: branch prediction validation happens in deep in the CPU pipeline. As a result, speculative instructions near the branch can access unprivileged memory locations.

- **P2**: speculative instructions leave side-effects in micro-architectural structures such as caches, which can be inferred using well-known timing side channel attacks like Flush+Reload [40] and Prime+Probe [29].

- **P3**: the branch predictor can be mistrained (Spectre 1), or directly polluted (Spectre 2). It is shared across all programs running on the same physical core [13, 23, 6], allowing code running in one privilege domain to manipulate branch prediction in another domain (e.g., kernel, VM, hypervisor, another process, or SGX enclave). Our attacks replace this step with speculation control through the RSB.

- **P4**: permission checks are performed deep in the pipeline and execution fault is generated only if the instruction is committed, enabling speculative instructions to access data outside its privilege domain;

```
if (offset < array1_size)
    y = array2[array1[offset] * 64];
```

Figure 1: Spectre attack variant 1

Spectre (Variant 1) is presented in Figure 1. In this code, a victim process reads values from `array1` using the `offset` provided by the attacker. Then, the resulting value is used to perform an access into `array2`. As we discussed above, accesses into the `array2` can be used by the attacker to deduce the value of the index. The index, in its turn, is controlled by the attacker since attacker controls the offset. Therefore, the attacker can use a carefully selected value of offset to read arbitrary memory address which then will result in cache access observable by the attacker. However, the if statement ensures there are no out of bounds memory accesses allowed. Unfortunately, the attacker can exploit speculative execution and behavior of branch predictor to force the victim process to perform an out of bounds memory access in the following way:

a) The attacker mistrains the branch predictor by executing the code several times with the value of the offset such that the if statement is true *(branch instruction not-taken)*.

b) Next, to make the speculative window larger, the attacker evicts array1_size from the cache, so that the CPU has to load the value from memory. Since the speculation result will not be resolved until this value arrives, forcing it to come from memory expands the size of the speculation window to allow more elaborate speculative gadgets to be executed.

c) Finally, the attacker chooses the malicious offset such that it be larger than array1_size. The trained branch predictor unit predicts the branch not-taken, so that the CPU executes two memory accesses speculatively and discloses the secret value through the cache side channel.

Prior work [6] had shown that the branch predictor is shared among processes on the same core. So, one thread can pollute it for another across protection boundaries (including across VMs). Thus, the attacker can poison the branch target predictor for the victim and force it to speculatively execute the gadget which reveals the sensitive data within the victim. This is a dangerous attack because it allows cross process/cross VM Spectre attacks.

The closely related Meltdown attack relies on the fact that a permission check for memory access during normal out-of-order execution of an instruction can happen late in the instruction execution due to pipelining and instruction reordering (P4) allowing the CPU to load the privileged data until the permission is later checked. Unlike Spectre variants, Meltdown does not rely on using misspeculation. Since an exception eventually will be raised, this attack requires the ability to tolerate and recover from the raised exception.

## 2.2 Defenses against Meltdown/Spectre

After the disclosure of Spectre and Meltdown in January, 2018 [13, 23, 27], a number of defenses were suggested.

**Intel proposed defenses:** Intel released a whitepaper [14] suggesting three types of defenses.

- To mitigate Spectre V1 attack, Intel recommends inserting a LFENCE instruction after the branch as a barrier to stop speculative execution. This defense mechanism has now been adopted by compilers such as GCC [30] and MSVC [32]. However, this does not prevent attacks where the attacker controls the program and does not use LFENCE instructions.

- To mitigate Spectre V2 attack, Intel introduced three new processor interfaces through microcode updates [16]:

  - Indirect Branch Restricted Speculation (**IBRS**) prevents software running in higher privileged mode from using prediction results from software running in lower privileged mode.

  - Single Thread Indirect Branch Predictors (**STIBP**) prevents code executing on one logical processor from impacting the indirect branch prediction of code executing on another logical processor.

  - Indirect Branch Predictor Barrier (**IBPB**) stops software running before the barrier from affecting the indirect branch prediction of software running after the barrier.

- To mitigate Meltdown, Intel recommends unmapping more privileged domain (kernel space) during the execution of less privileged software, which has been adopted by all popular operating systems, including Windows, Linux, and macOS. This is the KPTI defense described below.

**Kernel Page-Table Isolation (KPTI)**: Gruss et. al [11] introduced a protection technique called KAISER to protect against side channel attacks bypassing kernel level address space randomization (KASLR) [12]. The protection is based on unmapping kernel pages while in user mode, and remapping them on a mode switch to the kernel. As a result, misspeculation from user code is not able to access kernel memory, preventing Meltdown. It has been reported that KPTI can introduce substantial performance overhead [10]. KPTI cannot prevent attacks within the same privilege mode (e.g., to access memory outside a sandbox) [2, 34].

**Return Trampoline (retpoline)**: retpoline [37] is a software-based mitigation technique against indirect branch target injection attack (i.e., Spectre V2). It "exploits" two properties of the branch target prediction engine: (1) when executing a ret instruction, the predictor will utilize the return stack buffer (RSB) instead of the BTB; and (2) RSB cannot be polluted by attackers. The retpoline technique essentially swaps indirect branches for returns and deliberately pollutes the RSB with a useless gadget to control speculative execution. Retpoline protection requires access to source code and recompilation.

**RSB refilling (also known as RSB stuffing)** [15]: on Intel's Core i7 processors starting from Skylake (which are called Skylake+), an underfill condition in the RSB where a return occurs when the RSB is empty causes the processor to speculate the return address through the branch predictor. Thus, defenses deployed to protect indirect branches against Spectre variant 2 fail in this situations since return instructions can cause a misspeculation through the branch predictor. To counter this situation, Skylake+ processors also implement RSB refilling (a software patch): every time there is a switch into the kernel, the RSB is intentionally filled with the address of a benign delay gadget (similar to Retpoline) to avoid the possibility of misspeculation. RSB refilling interferes with SpectreRSB, although it was designed for a completely different purpose. However, we note that all Core i7 processors prior to Skylake are not patched with RSB refilling and that different processor lines, importantly including the Intel Xeon which are the primary platform used on Intel-based cloud computing systems and servers, are also unpatched, leaving them vulnerable to SpectreRSB.

# 3 Attack Principles: Reverse Engineering the Return Stack Buffer

In this section, we explain the operation of the Return Stack Buffer (RSB), which is the microprocessor structure our attacks exploit to implement speculation attacks that bypass all existing defenses. On modern processors, sophisticated branch predictors are used to predict the direction and target of conditional and indirect branches and calls. Return instructions challenge such predictors because the return address depends on the call location from which a function invoked, which for many functions that are called from different locations of a program can lead to poor branch predictor performance. For example, consider a function such as `printf()` which may be called from many different locations of a program. Relying on the previous history of where it returned to can lead to very low prediction performance through the branch predictor. We verify each of these mechanisms on two Intel processors (a Haswell and a Skylake).

## 3.1 RSB Overview

To overcome this problem, the return address is predicted using the RSB as follows. The RSB is a hardware stack buffer where the processor pushes the return addresses every time a call instruction is executed and uses that as a return target prediction when the matching return is encountered. Figure 2a shows an example of the state of the RSB after two function calls (F1 and F2) have been executed. The figure also shows the state of the software stack for the program where the stack frame information and the return address of the function are stored. Figure 2b shows how the values on these stacks are used when the return instruction from function F2 is executed. At this point, the return address from the fast shadow stack is used to speculate about the return address location quickly. The instructions executed at this point are considered speculative. Meanwhile, the return address is fetched from the software stack as part of the teardown of the function frame. The return address is potentially in main memory (not cached) and is received several hundred cycles later. Once the return address from the software stack is resolved, the result of the speculation is determined: if it matches the value from the RSB, the speculated instructions can be committed. If it does not, then a misspeculation has occurred and the speculatively executed instructions must be squashed. This behavior is similar to speculation through the branch predictor, except it is triggered by return instructions. Note that the misspeculation window could be substantially larger since the return could be issued out of order, and other dependencies have to be resolved before it is committed.

## 3.2 RSB sources of misspeculation

The RSB misspeculates when the return address value in the RSB does not match the return address value in the software stack, leading the program to misspeculate to the address in the RSB. If this misspeculation can be triggered intentionally by an attacker, spectre like attacks become possible through the RSB. Thus, in this subsection, we explain the sources of misspeculation through the RSB, and discuss whether they provide a vector for attackers to trigger speculation attacks. We label these sources as S1 to S4 to be able to refer to them in the attack descriptions.

**S1: Overfill or Underfill of the RSB due to limited structure size:** The RSB structure is typically sized to match common nesting depths of call stacks in programs. On low-end machines, the RSB can be as shallow as 4 entries in size. More typically, on desktops, it is in the range of 16 entries, and for server class processors, it can be larger (e.g., 24 entries on the AMD Ryzen [8]). As illustrated in Figure 3, when the RSB overfills, it typi-

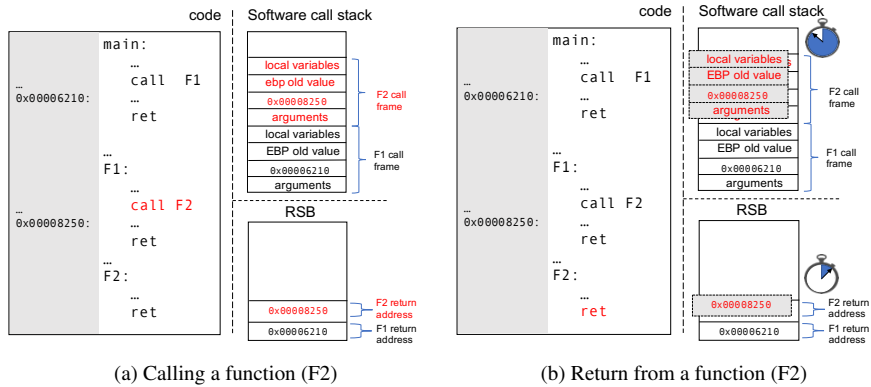

(a) Calling a function (F2)  (b) Return from a function (F2)

Figure 2: Example of function call and return effect on software call stack and RSB

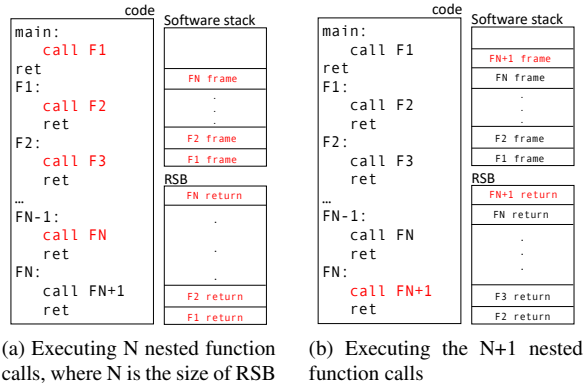

(a) Executing N nested function calls, where N is the size of RSB
(b) Executing the N+1 nested function calls

Figure 3: Example of overfill of RSB

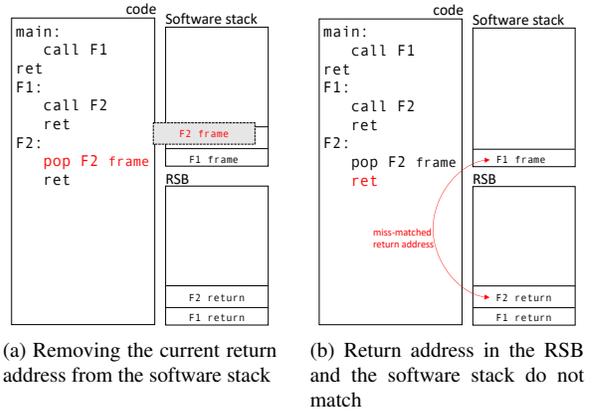

(a) Removing the current return address from the software stack
(b) Return address in the RSB and the software stack do not match

Figure 4: Example of direct pollution of the RSB

cally overwrites the older entries in the stack. Eventually, when the stack is unrolled as the nested calls return, we reach the function whose value has been overwritten causing an underfill of the stack (in Figure 3, the entry for F1 got overwritten).

In an underfill, there is no value available on the RSB to guide speculation. Different CPUs handle this situation differently. For example, the Intel CPUs that we checked switch over to the branch predictor if the RSB is empty, which can be used to trigger attacks through the branch predictor [25]. However, AMD appears not to follow this strategy.

**S2: Direct pollution of the RSB:** This is the primary vector that we use in our proof of concept attacks. Call instructions implicitly push a return address to the RSB and the software stack. However, an attacker can then replace the address on the software stack (by writing directly to that location), or just remove it altogether (as shown in Figure 4a). In this case, the value in the RSB remains and does not match a value on the software stack causing misspeculation when a return is executed (as shown in Figure 4b). By controlling the call address, the attacker can control the misspeculation address.

It is also possible to convert a call instruction into a push and jmp, in which case a return value exists on the software stack that is not matched by a value in the RSB. A return could also be replaced by a pop and a jump, causing a value to remain in the RSB that has been removed from the software stack.

**S3: Speculative pollution of the RSB:** speculatively executed calls push a value on the stack, although the details are specific to the architecture. Once misspeculation is discovered the call is squashed but the speculatively pushed return address remains on the RSB. This provides the opportunity for a malicious attacker to push a return address that is outside the address space accessible by the program (e.g., a kernel address) without raising an exception or having to handle the side effects of a call. [1]

---

[1] We did not use this vector, but its conceivable to use it to bypass Supervisor Mode Execution Prevention (SMEP) [9] to jump to a kernel gadget. For example, the user may attempt to jump to user code in PhysMap implementing a ret2dir [21] rather than a ret2usr [20] attack; the difficulty is that the user cannot pollute the RSB with kernel

**S4: RSB use across execution contexts:** on a context switch the RSB values left over from an executing thread are reused by the next thread. Once we switch to a new thread, if the thread executes a return, then it will misspeculate to an address provided by the original thread. The same is true with a switch over to the Operating System (provided RSB refilling is not implemented), or to an SGX context.

## 4 SpectreRSB: Basic attack example

In this section, we illustrate the attack principles by showing a basic speculation attack launched from a process to part of its address space that it cannot directly access (similar to Spectre variant 1 [13]). This attack represents the simplest instance of SpectreRSB and therefore we use it to explain the attack in detail. It is unlikely to be practical: it is difficult to implement the gadget to manipulate the stack using high level sandboxing primitives to allow the attack to break sandbox boundaries. On an unpatched machine, this attack enables the attacker to read kernel memory via the Meltdown bug. However, KPTI prevents using it to allow user code to read kernel data. We note that this attack does not rely on any speculation through branches or the branch predictor. For this reason, the attack bypasses defenses that focus only on securing speculation through the branch predictor. Most of our experiments were conducted on the machines shown in Table 1; the i7-6700 machine is a Core i7 Skylake with SGX2.

Figure 5 presents an overview of a basic SpectreRSB attack. The attack starts at line 22 with the call to `speculative`, with an argument which is the memory address of the sensitive data to be read. `speculative` calls `gadget`, which serves two purposes: (1) the return address is pushed to the RSB (the return address is to line 17 where we have the payload gadget to be executed speculatively); and (2) we jump to the (inline assembly) function `gadget` which will manipulate the software stack to create the mismatch between the RSB and the software stack. In this case, gadget cleans up the effects of the function call to itself, popping off the frame including the return address.

At this point, before the return, the stack state is consistent with a return from speculative back to main. However, the RSB holds a return value from `gadget` to `speculative`. Thus, in line 12 when the return executes, the CPU speculatively executes at line 17. The flush of the top of the stack (line 10) ensures that the true value of the return address will be fetched from memory rather than from the caches creating a large specu-

---

addresses using S2 without raising an exception, but may be able to do that using S3 (we note that PhysMap is marked as non-executable in most recent Linux distributions.)

```
1.  Function gadget()
2.  {
3.      push %rbp
4.      mov %rsp, %rbp
5.      pop %rdi    //remove frame/return address
6.      pop %rdi    //from stack stopping at
7.      pop %rdi    //next return address
8.      nop
9.      pop %rbp
10.     clflush (%rsp)  //flush the return address
11.     cpuid
12.     retq    //triggers speculative return to 17
13. }           //committed return goes to 23
14. Function speculative(char *secret_ptr)
15. {
16.     gadget();       //modify the Software stack
17.     secret = *secret_ptr;   //Speculative return here
18.     temp &= Array[secret * 256];    //Access Array
19. }
20. Function main()
21. {
22.     speculative(secret_address);
23.     for (i = 1 to 256)  //Actual return to here
24.     {
25.         t1 = rdtscp();
26.         junk = Array[i * 256];  //check cache hit
27.         t2 = rdtscp();
28.     }
29. }
```

Figure 5: SpectreRSB basic attack example

lation window. Note that the speculation window based on the return. Speculative execution at line 17 reads the secret which can be any mapped address even if inaccessible to the user process during normal execution and then communicates it out through the flush reload cache side channel by accessing a data dependent index in the Array (line 18). Finally, the real return value is obtained, and the misspeculation is squashed, returning us to line 23, where we probe the cache to identify which data dependent cache set was accessed to expose the value of the secret.

## 5 Attacks across different threads/processes

In this section, we investigate different vectors of SpectreRSB which exploit S4 (RSB use across execution context) to pollute the RSB. These attacks potentially allow an attacker to attack another process (Similar to Spectre V2), perhaps even across VMs, making the attack dangerous on the cloud. In general, these attacks require a machine not implementing RSB refilling (pre-Skylake, or Xeon, for example), to make sure that a context switch does not overwrite the polluted addresses from the RSB (Figure 6).

The attacker establishes co-location with the victim on the same core similar to Spectre 2. The attack pat-tern proceeds as follows. (1) after a context switch to the attacker, s/he flushes shared address entries (for flush reload). The attacker also pollutes the RSB with the target address of a payload gadget in the victim's address space; (2) the attacker yields the CPU to the victim; (3) The victim eventually executes a return, causing speculative execution at the address on the RSB that was injected by the attacker. Steps 4 and 5 switch back to the attacker to measure the leakage.

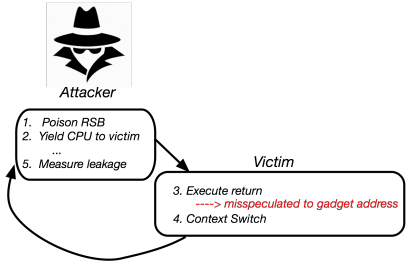

Figure 6: Attack 2: Basic Attack Flow

## 5.1 Attack 2a: Attack across two colluding threads

In this attack, the attacker and the victim are two colluding threads following the steps in Figure 6. In the first attack, we let the two threads synchronize using `futex` operations to control their interleaving. The RSB pollution happens in the first thread which also flushes the top of the stack of the second thread, while the return happens in the second. The attack succeeded, proving that SpectreRSB works from one thread to another. However, since the return is in user mode, we cannot read kernel data. For the attack to be useful, we should either launch an attack such that the victim colluding thread returns while in the kernel (enabling us to read kernel data while its memory is mapped), or work across process boundaries such that the victim thread is a different process and we leak its sensitive data (attack 2c).

## 5.2 Attack 2b: Attack with two colluding threads with return from inside kernel

Next, we wanted to see if we can use this attack to cause a return while the victim thread is in the kernel mode in step 3. To ensure this, we have the colluding victim execute a blocking system call, which typically has them deep inside a call stack in the kernel before blocking. The attacker after polluting the RSB, waits for the victim to unblock, perhaps even triggering the event that unblocks it. At this point the victim continues execution inside the kernel, and recurses back out of its call stack, with one or more returns, triggering the vulnerability. This attack requires a machine without SMEP enabled. We demonstrated the attack with SMEP disabled.

## 5.3 Attack 2c: Attacks across Process boundary

The attacks above assume two colluding threads. In principle, it can be generalized across different address spaces, but this requires overcoming some challenges. First, the attacker has to be able to identify gadgets existing in the victim binary instead of being able to use their own. This may also require them to recover the ASLR offset of the victim, but there are a number of existing attacks that make that possible. However, once these gadgets are found, the same attack pattern can follow by first polluting the RSB, then using eviction to remove the top of the stack containing the return address from the cache to extend the speculation window. Synchronization is difficult, but can be simplified if the attacker is able to trigger operations in the victim (e.g., if the victim is a server accepting connections). We did not create a PoC of this attack.

# 6 SpectreRSB Atack 3: Attacks on SGX

Having established attack 1 of the SpectreRSB where the attacker pollutes the RSB for its own process to cause misspeculation, we next investigate whether SpectreRSB attacks work on SGX compartments (similar to SGXSpectre [3]).

In this attack we consider, a malicious untrusted user code manipulates the RSB to try to cause misspeculation inside the enclave. In this attack, we pollute the RSB with the target address of a payload gadget from untrusted user code (this can equally be done by a malicious OS). Note that the gadget could be in the untrusted user code since user code and SGX enclaves share the same address space. The next step is to do an enclave call to switch it the trusted execution mode. The enclave call has to have an unmatched return to cause speculation execution at the address that was injected from the untrusted code. Finally, the untrusted code after returning from the enclave call can check the cache to record the leakage.

**Triggering an unmatched return:** the RSB assumes that strictly paired call-return behavior. In attacks that cross execution boundaries, the attacker pollutes the RSB, but would then like to trigger a return in the victim process code (or OS/SGX code) to which they have no access. However, if the attacker manages to catch the victim inside of a function call, then when the victim executes again, it will encounter an unmatched return. This could rely on timing or a blocking call inside of a function that will cause the scheduler to unschedule the victim. In this Proof of Concept attack, we placed an unmatched return directly in the enclave, but we expect to be able to do that using the strategies above for other enclaves.

**Attack results:** This attack successfully works on fully patched machines. The attack bypasses all software and microcode patches:it bypasses Retpoline since no indirect jumps are used. It bypasses the microcode patches since they do not appear to limit speculation through the RSB. It bypasses RSB refilling (which is only implemented on Skylake+, but not on the Xeon processors) since no mode switches to the kernel are triggered during the attack. Thus, SGX is vulnerable to SpectreRSB even on fully patched machines.

## 7 Potential Attack 4: From user to Kernel

In this section, we briefly discuss the possibility of another attack where user code pollutes the RSB and then triggers an unmatched return in the kernel (we call this attack 4). This attack is likely to be difficult, if not impossible, so we describe it only for completeness. The main insight is that a return from the kernel to a polluted address in the RSB will cause speculation while in kernel mode. This means that the kernel address space is still mapped, allowing us to read from kernel. This attack assumes the following ingredients: (1) that the RSB is shared between the user and the kernel: we find that this is the case on two Intel processors; (2) We need to be able to trigger an unmatched return in the kernel. Although some programming constructs such as tail recursion, continuations, setjmp/longjmp and others can break call-return semantics, we have not attempted to find such unmatched returns in the kernel; and (3) We need to figure out the stack address of the kernel, and evict it from the cache. This last step is necessary to make sure that the speculation window is sufficiently large to execute a useful gadget speculatively (without this, we can only execute a gadget a few instructions long speculatively). Luckily, the mapping between the stack kernel address and the physical address is deterministic in Linux on x86-64 (it uses the Physmap address directly instead of double mapping it). This makes deriving the conflict set straightforward once we identify the kernel stack address.

We explore a proof of concept attack with an unmatched return in a kernel module that we build. Later, we discuss concrete possibilities for how to make this happen with multiple threads. The attack is shown in Figure 7, and works only on an unpatched machine or a machine not implementing RSB refilling. After polluting the RSB in steps 1-3, and flushing the top of the kernel stack in step 4, before issuing a system call to our kernel module with the unmatched return . The mis-matched return triggers a misspeculation to step 7 to execute in supervisor mode. This attack does not work on patched Skylake+ processors due to RSB refilling but works on the Xeon machine. We also discover that Supervisor Mode Execution Prevention (SMEP) checks are not speculatively bypassed since the speculative program counter is known at the time of speculation. Thus, the attack as shown requires SMEP to be disabled to enable the kernel to return to user code. An alternative strategy to bypass this limitation is to try to use the return address of the gadget in PhysMap (as discussed under S3 in Section 3), but most Linux distributions disable execution of PhysMap addresses. We only demonstrated the attack with SMEP disabled. We also assumed that we know the address of the kernel stack pointer to flush it in step 4.

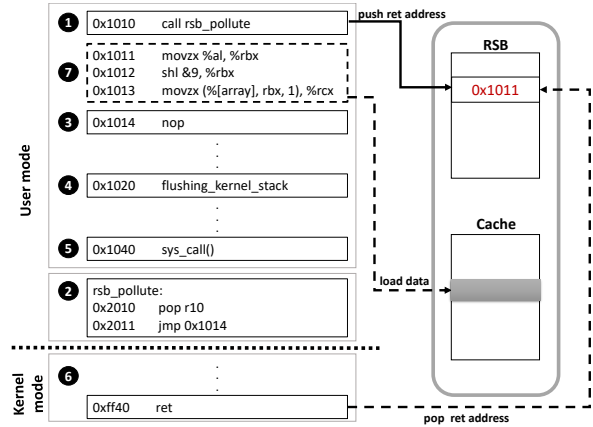

Figure 7: Attack 4: Basic Attack Flow

## 8 Discussion and Mitigations

In the first attack, a process launches an attack within the same address space either on the kernel or on data outside of its software containment. An attack from user mode to the kernel is possible in the original spectre: the BTB is poisoned, and then an indirect jump in the kernel space triggers the misspeculation to the payload. This attack is prevented by the Retpoline defense which only covers BTB poisoning, and assumes that the user code is compiled to use Retpoline. Importantly, none of the Intel microcode patches seem to restrict speculation through the RSB.

In attack 2 of SpectreRSB, we attempted to carry out the attack across execution threads. Attack 2a demonstrates a practical attack across two colluding threads in the same process. Attack 2b, shows how two colluding threads can cooperate to make an attack like attack 4 more predictably execute a return in kernel mode. Finally, Attack 2c in principle can carry out a SpectreRSB attack across different processes, bypassing all known defenses. Although we did not demonstrate this attack completely yet, we believe that none of the defenses stop

Table 1: Experiment Environment

|  | CPU Model | Kernel version | kernel patch | Intel patch |
|---|---|---|---|---|
| Machine1 | Intel Xeon(R)-E51620 | 4.15.0-22-generic | Retpoline, Kpti | ✓ |
| Machine2 | Intel Core(TM)-i7-6700 | 4.4.117 | Retpoline, Kpti, RSB refilling | ✓ |

Table 2: Attack senarios vs. defense mechanisms (✓attack can bypass defence or ✗otherwise). KPTI prevents reads from user mode to kernel memory but not reads from user to user or from kernel to kernel.

| Attack no | Attack Name | lfence | IBRS | STUBP | IBPB | retpoline | RSB refilling | SMEP/SMAP |
|---|---|---|---|---|---|---|---|---|
| Attack 1 | Same-process | ✓ | ✓ | ✓ | ✓ | ✓ | ✓ | ✓ |
| Attack 2a | Colluding threads (user) | ✓ | ✓ | ✓ | ✓ | ✓ | ✗ | ✓ |
| Attack 2b | Colluding threads (kernel) | ✓ | ✓ | ✓ | ✓ | ✓ | ✗ | ✗ |
| Attack 2c | Cross-process | ✓ | ✓ | ✓ | ✓ | ✓ | ✗ | ✓ |
| Attack 3 | Return in SGX | ✓ | ✓ | ✓ | ✓ | ✓ | ✓ | ✓ |
| Attack 4 | Kernel from user | ✓ | ✓ | ✓ | ✓ | ✓ | ✗ | ✗ |

it and it should be considered a dangerous threat vector on the cloud.

The attack on SGX bypasses all known defenses, including RSB refilling, and should be considered a dangerous open vulnerability on SGX systems. Changing the microcode patches to protect speculation through the RSB, or implementing RSB refilling upon entrance to SGX enclaves can potentially mitigate this vulnerability.

In Attack 4, our intuition was that a SpectreRSB attack that causes a return in the kernel to misspeculate would defeat both KPTI (in kernel mode, the kernel pages are available) and Retpoline (which only protects indirect jumps and calls, but not returns). While this is generally true, we discovered a number of complications that can be overcome under some conditions. RSB refilling, which is implemented on Intel Skylake+ processors, stops the attack. On other processors, SMEP prevents a return to a gadget in user space, however, a return to a gadget in kernel space if one can be identified is possible. Finally, we need to be able to determine the address of the top of the kernel stack in order to be able to evict, to increase the speculation window.

To mitigate SpectreRSB, we suggest that all processors, not just Skylake+ immediately support the RSB refilling patch which should interfere with all attacks that require a context switch to the kernel (attacks 3 and 4). Adding RSB refilling on an SGX enclave entrance should also be considered to stop attack 2. We also suggest that Intel microcode patches consider extending protection to the RSB, and not just the branch predictor. We summarize the proposed SpectreRSB attacks as well as their ability to bypass defenses in Table 2.

## 9 Related Work

In addition to the defenses against Meltdown and Spectre discussed in Section 2, we overview other potential defenses, as well as newly published related attacks. We first overview some of the defenses against side channel attacks on CPU caches, which are the primary channel used by the Meltdown and Spectre attacks to communicate privileged data out. We show that due to the fact that their threat model assumptions do not hold in speculation attacks, they fail to mitigate these attacks. Moreover, although making the cache secure protects the published proof of concept attacks, other side channels exist and it is straightforward to switch to them to communicate speculative data. Thus, a principled solution against speculation attacks should not rely on closing any particular side channel.

Software defenses against side channel attacks assume that the programmer and/or compiler are interested in preventing leakage from their program. In contrast, part of our threat model is an attacker that either writes the code to generate leakage, or identifies potential unintended code gadgets to cause the leakage. Thus, we believe that software side channel attacks are unlikely to provide beneficial defenses against speculation attacks.

**Hardware assisted cache side channel defenses:** Since speculation attacks rely on a microarchitectural covert channel in the payload gadget to communicate the data out, defenses against side channel attacks are a potential approach to mitigate speculation attacks. One simple technique to make caches immune to side channel attacks is static partitioning to create isolation [5]. Domnitser et al. propose a dynamically partitioned cache, providing isolation to limit leakage but also allowing some contention for performance [4]. We note that partitioning does not prevent leakage from within the same process such as the attacks under Spectre variant 1 and Meltdown. Locking of critical data [38, 24, 28] in the cache prevents it from being replaced by the attacker's prime or flush operations. The solution requires support from the OS, programming language, and compiler to mark the critical data, in addition to a bit for each cache line to indicate whether it is locked. The notion of sensitive data does not exist in the context of speculative attacks unless all out of bounds data is marked as sensitive (which would be impossible to fit in the caches). Randomization, exemplified by NewCache [39], randomizes the victim selection process on cache replacements, so that the attacker cannot glean useful information from its cache misses. The solution requires an index remapping table, extra bits in the cache to indicate which lines are subject to the random victim selection, and also support from the software layers to mark such critical data [39]. Kayaalp et al. [19] propose a defense that relaxes the inclusion property, which is a necessary component of the LLC attacks. Again, since an attack can be launched within the same process, neither randomization nor relaxed inclusion help with attacks within the same process (such as SpectreRSB attack 1 and 3, or the Meltdown bug).

**Hardware Solutions:** It is almost certain that future generations of CPUs will be designed to mitigate Meltdown and Spectre class attacks. To protect against Meltdown, it is possible to move the permission checks earlier in the pipeline, preventing the temporary load of the secret data. It is likely that AMD and ARM already implement this defense. Protecting against Spectre (including SpectreRSB) is substantially more difficult. To this end, SafeSpec [22] proposes using shadow hardware structures where speculative data resides. If the instructions fetching the data commit, the data is moved from these structures to the permanent structures (e.g., caches or TLBs). On the other hand, if an instruction is squashed, the data is discarded from the shadow structures. As a result, speculatively accessed data is never visible to committed instructions. SafeSpec requires additional space to store the shadow state, but results in modest improvements in the processor performance, while completely closing this class of vulnerabilities. PoisonIvy [26] is an architectural solution to track speculative data and prevent it from being exposed outside of the chip. The threat model focuses on accessing data while speculating on integrity verification. They seek to prevent the data from being speculatively read and therefore observed by a physical attacker that monitors the memory bus. PoisonIvy supports this capability by using information flow tracking to track data that is generated past a speculative check or data that is dependent on it. PoisonIvy does not prevent side channel leakage from speculatively accessed data. PoisonIvy results in approximately 20% slowdown in CPU performance. Both of these proposals require deep redesign of the processor architecture and therefore cannot protect current systems.

**Other attacks:** Since the disclosure of the Spectre/Meltdown attacks, two closely relevant attacks have also been reported [36, 3]. Utilizing a verification tool, Trippel et.al. [36] discovered that by leveraging the invalidation message of cache coherence protocols, it is possible to replace Flush+Reload with Prime+Probe to retrieve the content fetched by speculative instructions. In the SGXPECTRE attack, Chen et al. [3] demonstrated that it is possible to steal secret information from an SGX enclave using Spectre attack principles. Evtyushkin et al. presented BranchScope, an attack that can pollute the direction predictor (rather than the target predictor) component of the branch predictor unit [7]. Branchscope is likely to be less potent than attacks that poison the branch target buffer since it only controls the binary prediction of branch taken or not taken. It has not been demonstrated to be useful in a speculation attack, although it is possible that it can be. Although not a speculative attack, Branch Shadowing [25] empties the RSB to force returns inside of an SGX compartment to use the branch predictor, leaving a side channel footprint (observable through a side channel attack on the branch target buffer) enabling the control flow to be tracked. Recently, a so called variant 4 of Spectre was disclosed [17] which uses speculative store bypass: a speculation technique where load instructions speculatively execute without checking the load store queue for a preceding store. This technique represents another trigger for speculation, but it is not clear whether it can be used in practical attacks yet.

## 10 Concluding Remarks

In this paper, we introduced a new type of speculation attacks (SpectreRSB) that is triggered by the Return Stack Buffer (RSB), rather than the branch predictor unit. The RSB is used to predict the address of return instructions. We demonstrated a number of vectors that allow an attacker to cause RSB misspeculation. Using these techniques, we construct a number of attack vectors including attacks within the same process, attacks on SGX enclaves, attacks on the kernel, and attacks across different threads and processes. SpectreRSB bypasses all published defenses against Spectre, making it a highly dangerous vulnerability.

Interestingly, there is a patch that was proposed to protect against the behavior of Intel Core i7 Skylake generation and newer processors called RSB refilling. RSB refilling interferes with SpectreRSB attacks that experience at least one mode switch from user to kernel. We recommend that this patch should be deployed immediately across all processor generations (and not just Skylake+). In the long run, we believe that these patches are ad hoc and that new attack vectors will continue to emerge. Current systems are fundamentally insecure unless speculation is disabled. However, we believe that it is possible to design future generations of CPUs that retain speculation but also close speculative leakage channels, for example by keeping speculative data in separate CPU structures than committed data.

## Acknowledgements

The authors would like to thank the reviewers, and especially the paper's shepherd Daniel Gruss, for their valuable comments and suggestions which substantially improved the paper. This paper was made possible by NPRP grant 8-1474-2-626 from the Qatar National Research Fund (a member of Qatar Foundation). The statements made herein are solely the responsibility of the authors.